\documentclass{aastex63}

\usepackage{amsmath}

\usepackage{xcolor}
\newcommand{\COMON}{\begin{color}{blue}}
\newcommand{\COMOFF}{\end{color}}

\submitjournal{ApJ}

\shorttitle{CME at PSP and STEREO-A}
\shortauthors{Winslow et al.}

\graphicspath{{./}{Figures/}}

\begin{document}

\title{First simultaneous in situ measurements of a coronal mass ejection \\
by Parker Solar Probe and STEREO-A}

\correspondingauthor{Reka M. Winslow}
\email{reka.winslow@unh.edu}

\author[0000-0002-9276-9487]{Reka M. Winslow}
\affiliation{Institute for the Study of Earth, Oceans, and Space, University of New Hampshire, Durham, NH, USA}

\author[0000-0002-1890-6156]{No\'e Lugaz}
\affiliation{Institute for the Study of Earth, Oceans, and Space, University of New Hampshire, Durham, NH, USA}

\author[0000-0002-5681-0526]{Camilla Scolini}
\affiliation{Institute for the Study of Earth, Oceans, and Space, University of New Hampshire, Durham, NH, USA}
\affiliation{University Corporation for Atmospheric Research, Boulder, CO, USA}

\author{Antoinette B. Galvin}
\affiliation{Institute for the Study of Earth, Oceans, and Space, University of New Hampshire, Durham, NH, USA}



\begin{abstract}

{We present the first PSP-observed CME that hits a second spacecraft before the end of the PSP encounter, providing an excellent opportunity to study short-term CME evolution.} The CME was launched from the Sun on 10 October 2019 and was measured in situ at PSP on 13 October 2019 and at STEREO-A on 14 October 2019. The small, but not insignificant, radial ($\sim 0.15$~au) and longitudinal ($\sim 8^{\circ}$) separation between PSP and STEREO-A at this time allows for observations of both {short-term radial evolution} as well as investigation of the global CME structure in longitude. Although initially a slow CME, magnetic field and plasma observations indicate that the CME drove a shock at STEREO-A and also exhibited an increasing speed profile through the CME (i.e. evidence for compression). We find that the presence of the shock and other compression signatures at 1~au are due to the CME having been overtaken and accelerated by a high speed solar wind stream (HSS). We estimate the minimum interaction time between the CME and the HSS to be $\sim 2.5$ days, indicating the interaction started well before the CME arrival at PSP and STEREO-A. Despite alterations of the CME by the HSS, we find that the CME magnetic field structure is similar between the vantage points, with overall the same flux rope classification and the same field distortions present. These observations are consistent with the fact that coherence in the magnetic structure is needed for steady and continued acceleration of the CME.

\end{abstract}

\keywords{}


\section{Introduction} 

The Parker Solar Probe mission \citep[PSP;][]{Fox2016} {provides a} significant opportunity to obtain multi-spacecraft measurements of coronal mass ejections (CMEs), with PSP providing measurements closer to the Sun than 1~au, {and {\it Wind}}, ACE or STEREO providing measurements near 1~au. Observing the same CME at two different locations in near-radial alignment allows us to investigate the evolution of its properties, as has been done using Helios data in the 1970s and 1980s \citep[]{Burlaga1981} and planetary mission data in more recent years \citep[]{Winslow2016, Good2018, Vrsnak2019, Davies2020, Lugaz2020, Salman2020}. {However, few CMEs have so far been observed in situ by PSP \citep[e.g.][]{Korreck2020, Lario2020}, and none that have been complemented by in situ observations at other spacecraft.}

Multi-point simultaneous observations of CMEs by spacecraft with relatively small radial separation, but not in perfect longitudinal alignment, carries the advantage of revealing the global structure of the CME, especially large-scale variations in longitude. This assumes that no significant radial evolution of the CME takes place between the observation points given the small radial separation. Past such studies have shown that CMEs can have global cross-sectional shapes from almost circular to significantly distended \citep{Kilpua2011,Farrugia2011}, that many CMEs cannot be explained in terms of a simple flux rope model \citep{Kilpua2011}, and that there exists a scale length of longitudinal magnetic coherence inside magnetic ejecta (ME) \citep{Lugaz2018}. 

On the other hand, in situ CME observations at large radial separations in near-radial alignment allow us to probe the same portion of a given CME structure at different evolutionary stages. Being less affected by the intrinsic spatial variability of CME structures than observations at larger angular separations, they provide unique {opportunities for investigating} two major science questions related to CME propagation: (1) how does the CME morphology change due to interaction with solar wind structures such as the heliospheric current sheet \citep[]{Winslow2016}, stream interaction regions \citep[]{AlShakarchi2018,Winslow2021}, or other CMEs \citep[]{Lugaz2017, Wang2018, Palmerio2019, Scolini2020}; and (2) how does the CME expand and its sheath region form \citep[]{Janvier2019, Lugaz2020, Salman2020}. 

Answering these questions also requires the consideration of the different aspects characterizing the propagation of slow and fast CMEs. In particular, the most studied CMEs tend to be large and fast events (the Halloween storm event, St. Patrick Day CMEs, the 2012 July 23 CME at STEREO-A, etc.). The physical processes affecting the propagation of slower CMEs have been investigated in far less detail. However, they may differ in essence from those occurring during the propagation of the best studied fast and wide events. For example, slow CMEs are much more likely to be overtaken by a fast solar wind stream rather than propagating through it; the shock is likely to form in the upper corona or even in the inner heliosphere rather than in the low corona; and the sheath may be a more complex combination of swept-up compressed and shocked solar wind \citep[]{Salman2020b}. Investigating these processes typically requires combining remote-sensing observations, numerical simulations as well as multi-spacecraft in situ measurements.

In this paper, we highlight {a} CME launched from the Sun on October 10, 2019, which was the first one to be observed simultaneously by PSP and another spacecraft in near-longitudinal conjunction. The separation between PSP and STEREO-A at this time was such that both small changes in radial evolution of the CME as well as some details about its global structure can be deduced. Although deemed a relatively ``poor" and slow event, this stealth CME's interaction with a high speed solar wind stream can improve our understanding of how typically understudied slow CMEs interact with large-scale structures in the solar wind. In Section~\ref{sec:arrival}, we present the in situ measurements of the CME at PSP and STEREO-A and provide clear evidence that the two spacecraft measured the same CME. In Section~\ref{sec:remote}, we determine the source region of this CME and information about the CME propagation in the inner heliosphere that can be inferred from the remote-sensing combined with the arrival times and speed at the two spacecraft. We compare the PSP and STEREO-A measurements and discuss the cause of measured differences in Section~\ref{sec:discussion}. We conclude in Section~\ref{sec:conclusion}. 

\section{CME Arrival and Characteristics at Parker Solar Probe and STEREO-A}\label{sec:arrival}

On 2019 October 13 at 19~UT, PSP was at $\sim$ 0.81~au, at $\sim 4^\circ$ latitude, and $\sim -75^\circ$ longitude in heliocentric Earth equatorial (HEEQ) coordinates. The spacecraft was moving anti-sunward at a speed of $\sim 0.01$~au per day, while its angular position with respect to the Sun--Earth line was essentially constant. STEREO-A was separated from PSP by about $\sim$0.15~au in radial distance, $< 1^\circ$ in latitude, and $\sim -7.7^\circ$ in longitude. The fairly close longitudinal alignment between PSP and STEREO-A allowed the CME in question to be observed at both spacecraft, while the proximity in radial distance made possible the first simultaneous observations of a CME by PSP and STEREO-A.

\begin{figure}
\centering
{\includegraphics[width=\hsize]{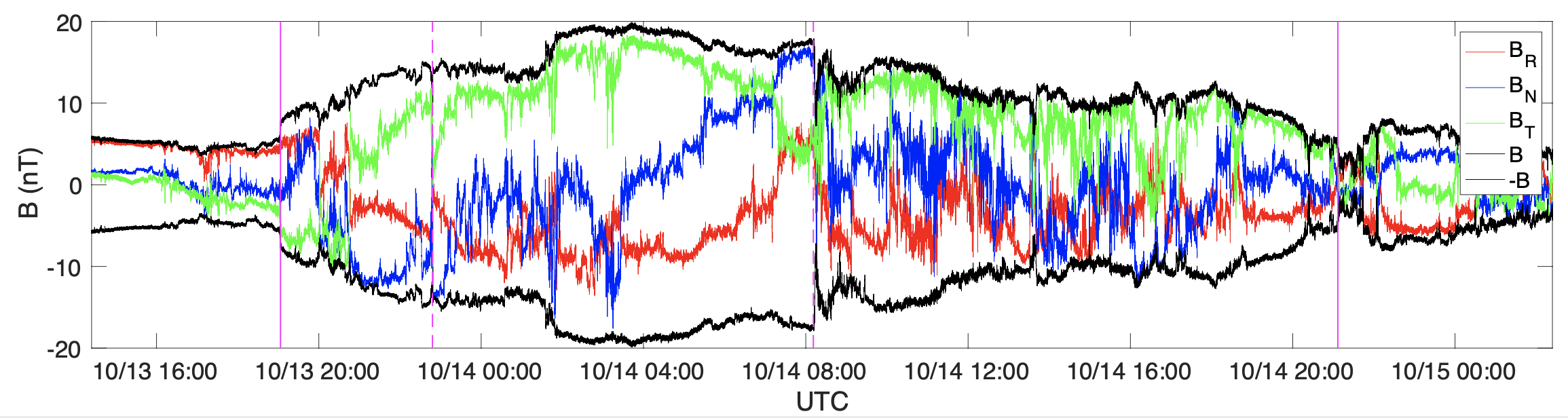}}
\caption{
Magnetic field data for the October 10, 2019 CME event, measured in situ at PSP on October 13--14, 2019. The data are in RTN coordinates with $B_R$ in red, $B_T$ in green, $B_N$ in blue. The magnetic field magnitude $B$ and its negative $-B$ are in black. The same color coding for the magnetic field data is used in the following figures as well. 
The vertical magenta lines denote the start times of the shock and the end time of the CME. 
The first dashed line marks the beginning of the ME, while the second dashed line marks the end time of the flux rope portion of the ME.
} 
\label{fig:CMEatPSP}
\end{figure}

Due to its proximity to aphelion, PSP was not taking plasma measurements during this time period. We rely therefore on magnetic field measurements by FIELDS \citep[]{Bale2016} to identify the arrival of various substructures within the CME. Figure~\ref{fig:CMEatPSP} shows the magnetic field measurements at the time of CME passage at PSP. There is a shock-like discontinuity that arrived on October 13, 2019 at 19:03~UT, as marked by the first solid vertical line. The arrival of the ME occurred at 22:48~UT, while the end of the CME was marked by a reverse shock on October 14, at 21:07~UT. 

\begin{figure}
\centering
{\includegraphics[width=\hsize]{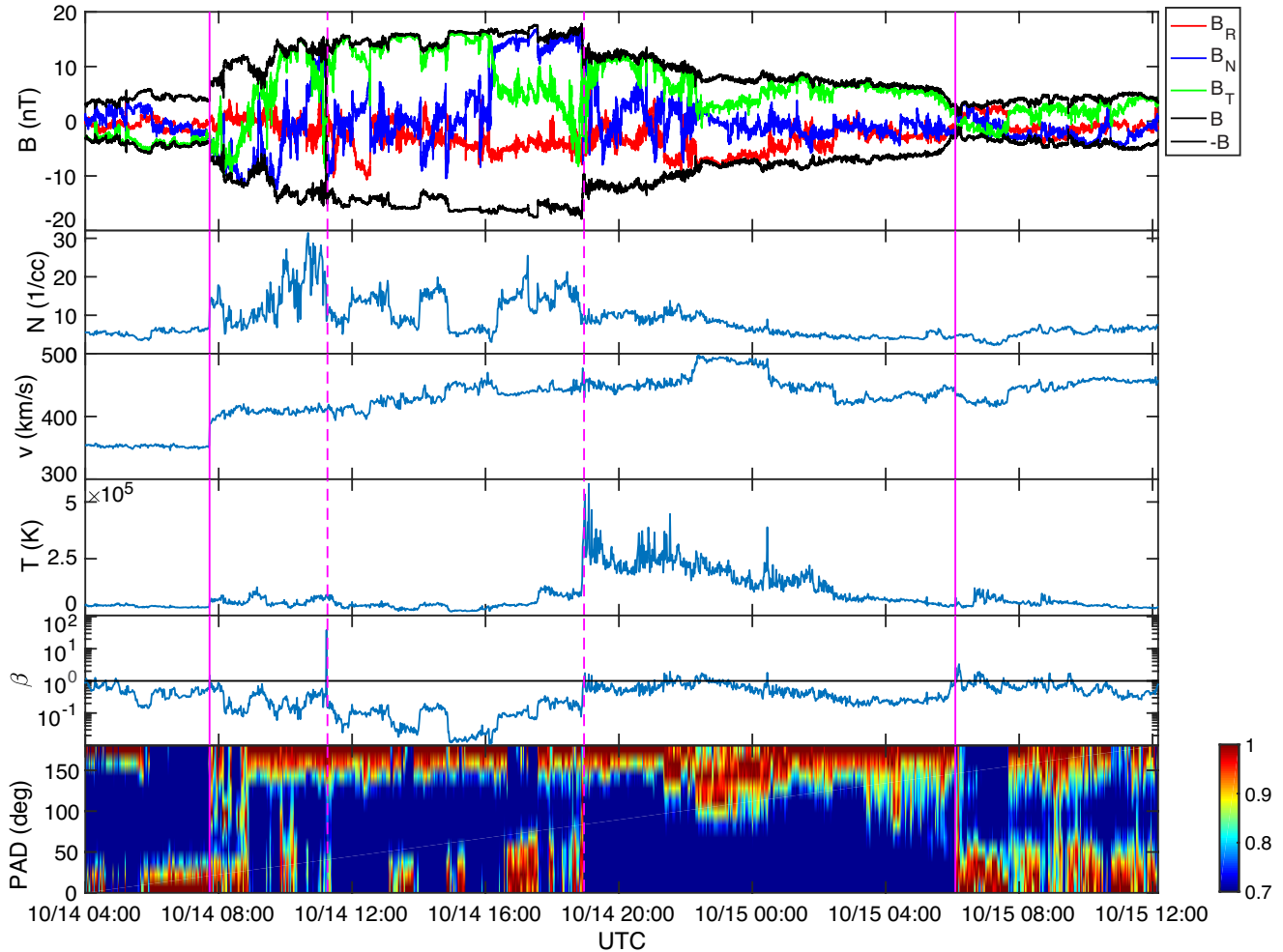}}
\caption{
Data for the October 10, 2019 CME event measured in situ at STEREO-A on October 14--15, 2019. 
The first panel shows the magnetic field data in RTN coordinates, using the same colors as in Figure~\ref{fig:CMEatPSP}. 
The second, third, fourth, and fifth panels show the proton number density, speed, temperature, and proton $\beta$, respectively.  The last panel shows the normalized suprathermal electron pitch angle distribution. The vertical magenta lines denote the start times of the shock and the end time of the CME. The first dashed line marks the beginning of the ME, while the second dashed line marks the end time of the flux rope portion of the ME.
} 
\label{fig:CMEatSTA}
\end{figure}

{We note here that we use the term ME in a similar manner to the way it was used in \citet{Winslow2015}. In cases such as this at PSP, when there are no plasma observations and therefore it cannot be verified whether the ICME also exhibited the expected plasma $\beta$ and temperature decrease associated with magnetic clouds (MCs), we find it prudent to use the term ME instead of MC. We thus refer to all potential MCs by the more generic term magnetic ejecta (ME) in this paper. We also note that the magnetic field inside the ME at PSP was unusually variable and disturbed compared to the cleaner, simpler MC-like events that are typically studied.} Although this made the identification of the ME start time at PSP particulary difficult, we were able to determine it by comparing the in situ observations at PSP with those at STEREO-A (Figure~\ref{fig:CMEatSTA}), where plasma data (which more clearly mark the ME boundaries) are available for this time period. Fortuitously, the same main features of the magnetic ejecta exist at PSP and at STEREO-A, including the discontinuity that marks the beginning of the ME, the sharp step-function-like drop in magnetic field magnitude that marks the end of the smooth field rotation (dashed vertical line on Figure~\ref{fig:CMEatPSP}), and a long, highly variable tail region of the ME. We thus used these magnetic field features identified at both spacecraft to determine the substructure boundaries at PSP (see Section~\ref{sec:discussion} for further details on the magnetic field comparison between PSP and STEREO-A). Given the especially high variability of the magnetic field in the tail of the ME at PSP, we have identified a region prior to it, within the ME, that exhibits flux rope-like signatures. This period of smoother variation of the magnetic field components with a rotation from $B_N$ negative to positive, with a predominantly positive $B_T$, occurred between the start of the ME and 14 October at 08:11~UT as marked by the dashed vertical lines. We also note that although suprathermal electron pitch angle data at PSP does exist at this time, it is of too low resolution to be useful for any magnetic boundary/connectivity analysis.

A CME with similar features (see also discussion in Section~\ref{sec:discussion}) arrived at STEREO-A starting with a fast-forward shock on 14 October 2019 at 07:44~UT. Solar wind plasma and magnetic field measurements are available from the STEREO PLASTIC \citep[]{Galvin2008} and IMPACT \citep[]{Luhmann2008} instrument suites, including suprathermal electron pitch-angle distributions. Measurements are shown in Figure~\ref{fig:CMEatSTA}, with the solid and dashed vertical lines corresponding to the same boundaries as those identified at PSP. Due to the availability of plasma measurements, the STEREO-A data gives us a clearer picture of some of the CME features already observed at PSP. After the shock, the ejecta starts on 14 October at 11:16~UT, marked by a similar discontinuity in the magnetic field measurements as at PSP along with a clear and sustained drop in plasma $\beta$, as well as a drop in proton density. There is also the same sharp drop in magnetic field magnitude as seen at PSP, observed at 18:57~UT on October 14 at STEREO-A, which is associated with a clear rise in the proton temperature and plasma $\beta$, the end of intermittent bidirectional electron signatures, as well as the beginning of the variable tail region. The data section between the start of the ME and the start of the ME tail is consistent with flux rope signatures, including intermittent bidirectional electrons, plasma $\beta$ well below 1, as well as decreased proton density and temperature. The magnetic field remains elevated until 06:05~UT on 15 October, which marks the end of the CME. The ME tail regions exhibits a dramatic rise in proton temperature, increase in plasma $\beta$ (although still below 1 on average), and a complete lack of bidirectional electrons in the suprathermal electron data. These signatures provide evidence that the ME tail underwent reconnection-related alterations. 

Furthermore, given that a very similar magnetic field structure is observed at PSP and at STEREO-A in the ME (both having a flux rope and separate tail section in the ME), we infer that the CME underwent alterations prior to reaching both PSP and STEREO-A. We also note a relatively unusual feature that the speed in the ME increases from the start of the ME until approximately half-way through the ME tail (see Figure~\ref{fig:CMEatSTA}). {It is more common to observe decreasing velocity within an ICME} because in most cases the CME is observed to expand, as indicated by a linear speed decrease through the ME, i.e., the front of the ME travels faster than the rear. Given the opposite scenario here, the fact that the rear of the ME travels faster than the front implies that the ME is shrinking, i.e., that it is being compressed. This compression is also clearly evidenced by the fact that the CME is of shorter duration at STEREO-A (22.4~hours) than at PSP (26.1~hours), {which is an overall duration difference over the entire ICME. We calculated the compression speed over the ME flux rope section only (where there is a clear linearly increasing speed profile) to be $\sim$20 km~s$^{-1}$, leading to an expected compression of the ME by $\sim$1.3 hr, in fairly good agreement with the observed 1.7 hr duration difference of this section between PSP and STEREO-A.} This is further discussed in Section 4.2. Finally, it is also important to note that although the arrival times of the CME substructures indicate simultaneous observations of the CME at PSP and STEREO-A, they are not simultaneous observations of the same features of the CME; PSP was immersed in the ME tail region when the CME arrived at STEREO-A. 

\section{Remote-sensing observations and CME kinematics}
\label{sec:remote}

%
\begin{figure}
\centering
{\includegraphics[width=\hsize]{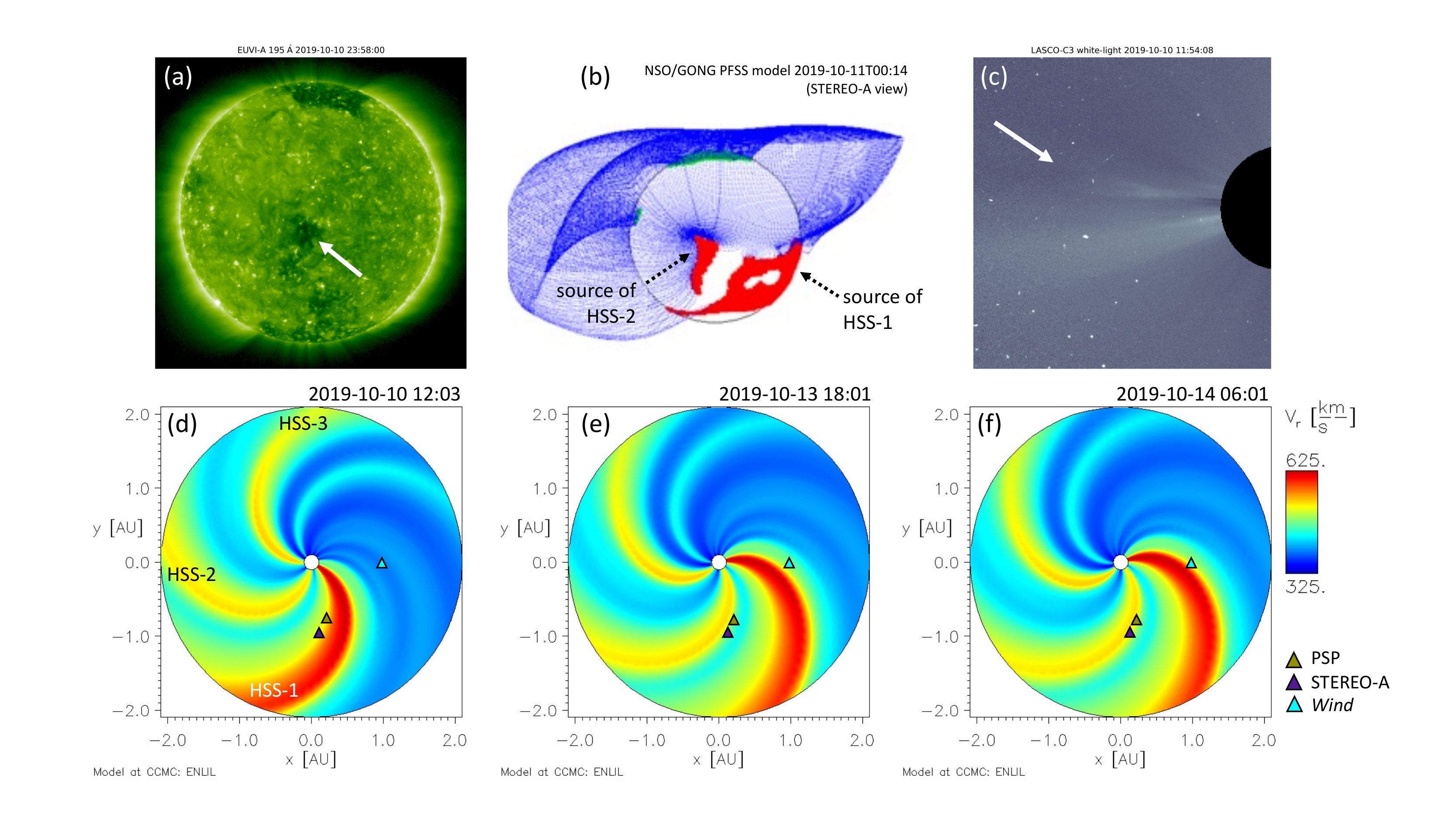}}
\caption{
Remote-sensing observations and heliospheric CME propagation as simulated by the ENLIL model.
(a) EUVI-A image in the 195 \AA \, filter on October 10, 00:58~UT. A coronal hole, probable source of HSS-2 observed at STEREO-A in the days following the passage of the CME (Figure~\ref{fig:CMEatSTA}), is indicated by the white arrow.
(b) NSO/GONG PFSS extrapolation around the same time (October 11, 00:14~UT), as seen from STEREO-A. {Two coronal holes, probable sources of HSS-1 and HSS-2 observed in situ at STEREO-A before and after the CME passage (Figure~\ref{fig:CMEatSTA_supplemental}), are indicated by black arrows.}
(c) LASCO C3 white-light image on October 10, 11:54~UT. The CME is highlighted by the white arrow.
(d), (e), (f): solar wind speed in the heliographic equatorial plane as modelled by ENLIL on October~10, 12:03~UT (while the CME was visible in LASCO~C3), on October 13,~18:01~UT (one hour before the CME arrives at PSP), and on October 14,~06:01~UT (less than two hours before the CME arrived at STEREO-A). {The three HSSs observed at STEREO-A during the period of interest (Figure~\ref{fig:CMEatSTA_supplemental}) are marked in panel (d).}} 
\label{fig:obs_mod_summary}
\end{figure}

\subsection{Remote-sensing observations}

The associated CME is listed in the CDAW LASCO CME catalog (\url{https://cdaw.gsfc.nasa.gov/CME_list/}) as first observed by LASCO~C2 on board SOHO on 10 October 2019 at 00:48~UT, and entered the C3 field of view around 02~UT. The event, propagating at a position angle of $82^\circ$ and with an angular width of $19^\circ$, is listed as ``very poor'' in the CDAW catalog, and it presents a clear accelerating behaviour while propagating through the C2 and C3 fields of view. The average projected speed is 180~km~s$^{-1}$, while the second-order speed at 20~solar radii is 282~km~s$^{-1}$.
Given that the CME was directed towards STEREO-A and PSP, we searched for additional coronal signatures of the event in COR2-A, but could not identify the CME in such images. This is probably due to a combination of its slow speed (which makes the CME faint), and narrow width (the CME might have been blocked by the occulting disk in COR2-A).
In the low corona, we report the lack of eruptive signatures in both SDO/AIA and EUVI-A images, suggesting this CME belongs to the class of ``stealth'' CMEs \citep[e.g.][]{Robbrecht2009, Nitta2017}.
Because of the single-viewpoint observations in the corona and lack of low-corona eruptive signatures, performing a {Graduated Cylindrical Shell (GSC) \citep[GCS;][]{Thernisien2006, Thernisien2009}} reconstruction is not possible. For this reason, we based the following analyses on the kinematic and geometric properties listed in the CDAW catalog, assuming that the CME propagates in the plane of the sky as seen from LASCO. Note that STEREO and PSP were located at $\phi = -83^\circ$ and $-75^\circ$, respectively, so that assuming the CME propagated in the plane of the sky implies a 7$^\circ$--to--15$^\circ$ angular separation between the CME direction of propagation and the spacecraft position, respectively.

Low-coronal EUV images and extrapolations of the coronal magnetic field topology using potential field source-surface models (Figure~\ref{fig:obs_mod_summary}(a), (b)) also indicate the presence of two coronal holes on the solar disk facing STEREO-A around the time of the CME eruption. A coronal hole characterized by an elongated and tilted morphology reaching equatorial latitudes with its northern portion is visible in the south-west quadrant of Figure~\ref{fig:obs_mod_summary}(b). A more compact, equatorial coronal hole crossing the central meridian around October 11, 00:00~UT as observed by STEREO-A is visible in Figure~\ref{fig:obs_mod_summary}(a) and (b). As further discussed in Section~\ref{subsec:backgrond_solar_wind_conditions}, these coronal holes are likely the sources of two HSSs arriving before (HSS-1) and immediately after (HSS-2) the CME at STEREO-A (Figure~\ref{fig:CMEatSTA}), suggesting the CME source region was located between the two of them.

\subsection{CME transit speeds}

We use the last CME height and time of observation as listed from the CDAW catalog (i.e. $r_0 = 6.95$ solar~radii at $t_0 = $ October 10 at 03:06~UT), together with the shock arrival times at PSP and STEREO-A, to compute the transit speeds of the CME between the Sun and PSP, the Sun and STEREO-A, and PSP and STEREO-A.
We find that the CME transit speed from $(r_0, t_0)$ to $(r_\mathrm{PSP}, t_\mathrm{PSP}) = (0.81$~au, October 13 at 19:03~UT$)$ is $363$~km~s$^{-1}$,
the CME transit speed from $(r_0, t_0)$ to $(r_\mathrm{STA}, t_\mathrm{STA}) = (0.96$~au, October 14 at 07:43~UT$)$ is $383$~km~s$^{-1}$,
and that the CME transit speed from $(r_\mathrm{PSP}, t_\mathrm{PSP})$ to $(r_\mathrm{STA}, t_\mathrm{STA})$ is $525$~km~s$^{-1}$.

The transit speeds from the Sun to the two spacecraft, along with the in situ measured speed at STEREO-A, are significantly higher than the initial CME speed as measured by LASCO/C3. This is consistent with this being a slow CME which accelerates as it propagates. The fact that the average transit speed is larger at STEREO-A than PSP implies that additional acceleration of the CME occurred between PSP and STEREO-A. Given the slow nature of this CME, one would expect it to accelerate until the solar wind speed is reached and then to continue at constant speed thereafter. However, the observations for this CME are consistent with a nearly linearly increasing speed from the Sun to 1 au, along with the evidence for acceleration in the speed profile of the CME at STEREO-A. These observations are consistent with an accelerating force at the back of the CME not being fully balanced by drag in the front of the CME as it is piling into the solar wind ahead of it. Further discussion of this is deferred to Sections~\ref{subsec:backgrond_solar_wind_conditions} and \ref{sec:conclusion} below.


\section{Discussion}\label{sec:discussion}

As PSP measurements only include reliable magnetic field measurements, we focus on the comparison of the magnetic field associated with the interplanetary shock, the sheath region, and the ME. 
\begin{figure}
\centering
{\includegraphics[width=0.8\hsize]{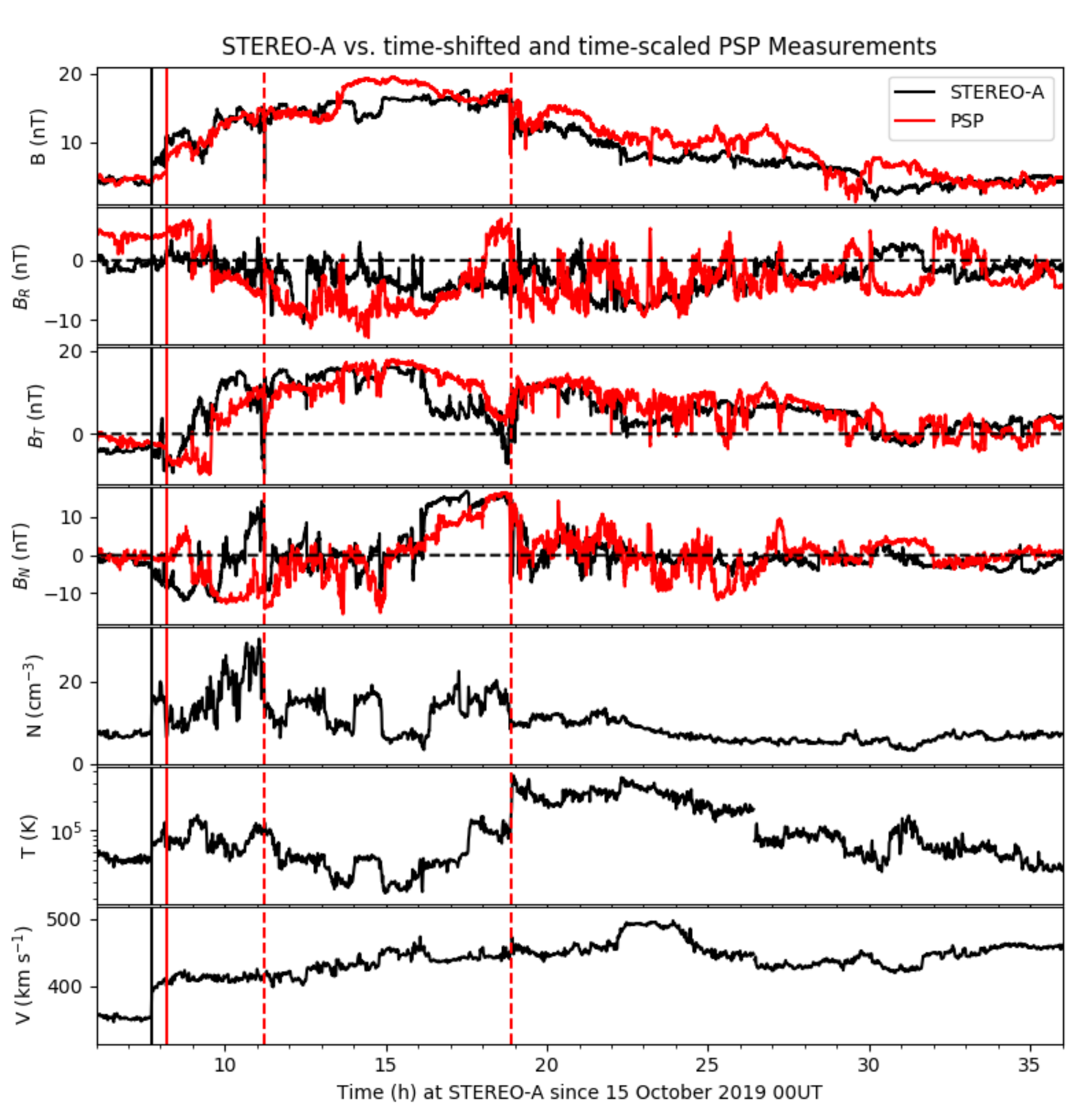}}
\caption{Overlay of the {\it in situ} measurements by STEREO-A (black) and PSP (red). The panels show, from top to bottom, the magnetic field strength, the radial ($B_R$), tangential ($B_T$) and normal ($B_N$) components of the magnetic field, and for STEREO-A only, the proton density, temperature and velocity. The PSP data are scaled and shifted for time only to obtain the same duration of the ME between the two dashed vertical red lines. The shock-like discontinuity at PSP is marked by the red  vertical solid line,  while the shock at STEREO-A is marked by the black  vertical solid line. Details of the time-scaling are given in the text. There is no scaling of the magnetic field strength or components.} 
\label{fig:STA_PSP_scaled}
\end{figure}

\subsection{Shock}
\label{subsec:shock}
At STEREO-A, there is a clear fast-forward shock at 07:44~UT on October 14. Before the shock, the Alfv{\'e}n speed is $\sim 30 \pm 2$~km~s$^{-1}$ and the sound speed is $\sim 20 \pm 2$~km~s$^{-1}$. To be consistent between PSP and STEREO-A measurements, we only calculate the shock normal using the magnetic coplanarity. At STEREO-A, we obtain the shock normal of $(0.92,-0.02 ,0.40)$ for a shock normal angle (angle between the shock normal and the upstream magnetic field) of $71 \pm 10^\circ$, corresponding to a quasi-perpendicular shock. The magnetic compression ratio is $1.75 \pm 0.15$. The error bars are estimated using slightly different intervals for the upstream and downstream regions. The jump in the normal velocity through the shock is $44 \pm 5$~km~s$^{-1}$. Due to the low upstream sound and Alfv{\'e}n speeds, this small jump in velocity is consistent with a weak shock of Alfv{\'e}n Mach number $M_A = 1.5 \pm 0.25$ or fast magnetosonic Mach number $M_{ms} = 1.4 \pm 0.3$. 

At PSP, there is a potential fast-forward shock or fast magnetosonic wave upstream of the magnetic ejecta clearly identifiable at 19:03~UT on October 13. However, the compression ratio is only  $1.35 \pm 0.1$. In addition, the upstream magnetic field strength is about 50$\%$ larger at PSP than it is at STEREO-A. Using the coplanarity relation, the normal to this discontinuity is $(0.93, 0.25, 0.25)$ consistent with that found at STEREO-A. This gives a shock normal angle of $\sim 55^\circ$. While there are no plasma measurements at PSP, a decrease of the density as $1/r^2$ between PSP and STEREO-A would be consistent with a slightly higher Alfv{\'e}n speed at PSP of $\sim 37 \pm 2$ km~s$^{-1}$. This would be consistent with a weaker shock or possibly a fast magnetosonic wave being in the process of sharpening into a shock. 

In light of the slow initial speed of the CME, the presence of a shock at STEREO-A and possibly at PSP is surprising. It is, however, consistent with the findings from the transit speeds (see Section~\ref{sec:remote}), which indicate that the CME must have accelerated significantly between the Sun and PSP/STEREO-A. This is also evident from comparing the CME initial speed with the in situ speed measured at STEREO-A. Interaction with a fast solar wind stream overtaking the CME is the likely cause of this acceleration (see Section~\ref{subsec:backgrond_solar_wind_conditions} for more detail). The fact that the CME drives a shock at STEREO-A is a clear indication that this interaction must have occurred for a relatively long time before the CME impacts PSP/STEREO-A. With the Alfv{\'e}n speed inside the ME at STEREO-A being $\sim 120-150$~km~s$^{-1}$ and the ME radial size being $\sim 0.19$~au at STEREO-A, the Alfv{\'e}n crossing time of the ME is $\sim 2.5$ days. This gives an estimate of the minimum time when interaction must have started at the back of the ME for the entire ME to have been affected and a shock to form.

\subsection{Correlation of PSP and STEREO-A Magnetic Field Measurements }
We follow the procedure of \citet{Lugaz2018} also used in \citet{AlaLahti2020} to compare the magnetic field measurements at PSP and STEREO-A. As a summary, we determine the time shift, component by component, that maximizes the correlation between the magnetic field measurements at STEREO-A and PSP. Here, STEREO-A and PSP are separated by $\sim 0.12$~au in the non-radial direction ($\sim 7.7^\circ$) and $\sim 0.15$~au in the radial direction, which is a similar ratio but about 10 times farther than the events possible from {\it Wind}--ACE conjunctions from the aforementioned studies. Separations of this magnitude have last been possible around 2007 when the two STEREO spacecraft were still close to Earth. The sheath region is of relatively short duration, and the two spacecraft have both radial and non-radial separations. The fact that the spacecraft have non-negligible radial separations as compared to these past studies implies that propagation effects may also affect the comparison of the measurements. For these reasons, we do not try to fit the sheath region separately from the ME. 

We obtain the best time-shifted correlation coefficients between the PSP and STEREO-A measurements for the entire ME (from the first dash line to the solid line indicating the end of the ME in Figure~\ref{fig:CMEatPSP}). For a time-shift of about 10.7~hours, the best correlation coefficient is 0.741, 0.325, 0.718, and 0.718 for the total magnetic field strength, the radial, tangential, and normal components, respectively. The value for the magnitude is consistent with the findings of \citet{Lugaz2018} even though the separation here is 10 times larger. This indicates that the correlation decreases by $\sim 0.3$ for every 0.1~au. The correlation for the $B_N$ and $B_T$ components is larger than some events reported in that study and indicates that the consistency (or coherence) of the magnetic field measurements inside MEs may be larger than previously reported.

Figure~\ref{fig:STA_PSP_scaled} shows an overlay of the PSP (red) and STEREO-A (black) measurements with a time-shift  and time-scaling to reflect both propagation (shift) and possible expansion/compression of the ME (scaling). It is important to note that no scaling was applied to the magnetic field components between PSP and STEREO-A in Figure~\ref{fig:STA_PSP_scaled}, only to the time component. A single value of the shift and scaling is used for all magnetic field components. The time for the PSP measurements is compressed by a factor of 1.235 
to match the duration of the part of the ME between the two dashed line in Figure~\ref{fig:CMEatPSP}, and a time-shift is applied to align the ME crossing at both spacecraft. The dashed lines correspond to clear discontinuities in the magnitude and tangential component of $B$, which allows us to easily find this best scaling to match the measurements. 

This Figure illustrates a number of similarities and differences between the CME at the two spacecraft: (i) consistent with the results of the correlation, the tangential and normal components of the magnetic field inside the ME show clear similarities even for an angular separation of more than 7$^\circ$, (ii) the ME has a longer duration and has somewhat higher peak magnetic field at PSP than at STEREO-A, (iii) there are significant differences in the normal component of the magnetic field inside the sheath region and the scaled sheath has a longer duration at STEREO-A than PSP. We also note that the fact that the peak field at PSP is higher than at STEREO-A is somewhat surprising given the observed compression in the CME {(based on the speed increase in the ME at STEREO-A and the duration decrease, as a proxy for size, between PSP and STEREO-A); one would expect the magnetic field to increase with compression.}

\subsection{Flux Rope Characterization}

\begin{figure}
\centering
{\includegraphics[width=\hsize]{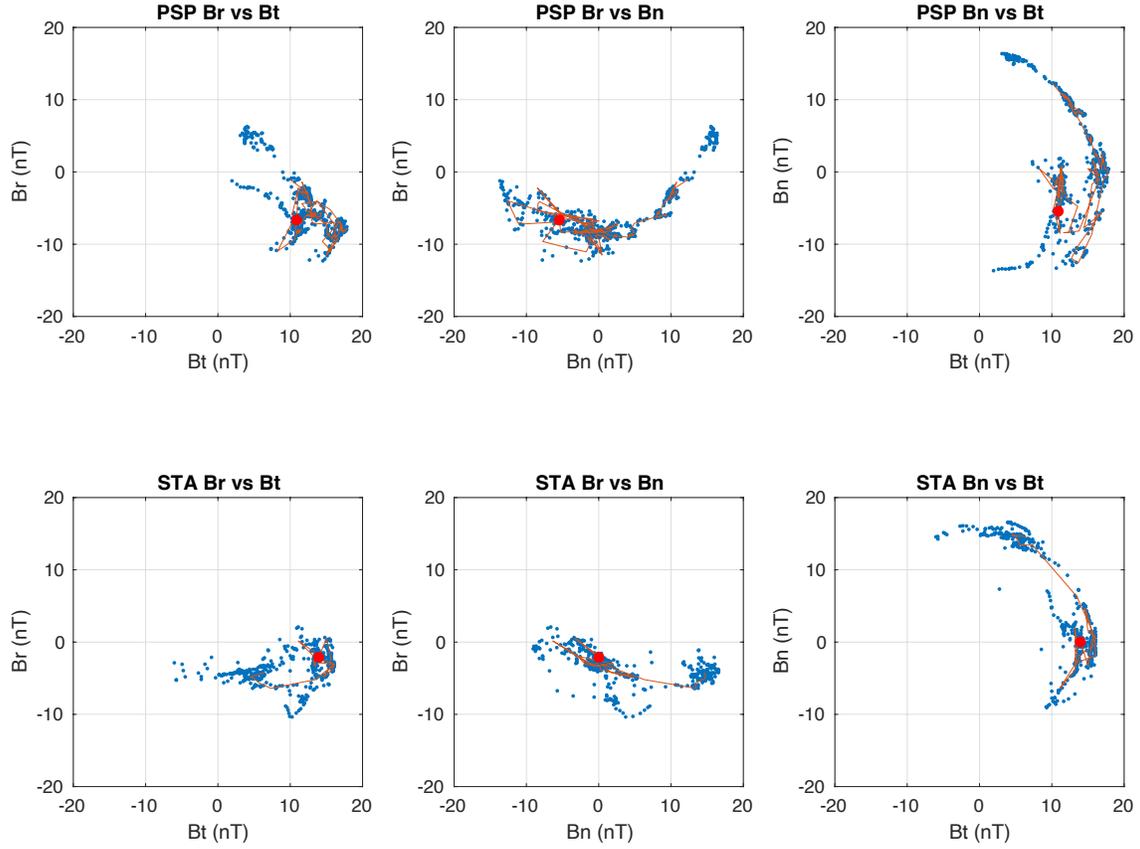}}
\caption{
Hodograms of the flux rope portion of the CME magnetic ejecta at PSP and STEREO-A in RTN coordinates. The red dots mark the measurements of the initial magnetic field components. 
} 
\label{fig:hodograms}
\end{figure}
Similarly to our methods in \citet{Winslow2021}, we use the classification method proposed by \citet{NievesChinchilla2019} to classify the CME flux rope structure based on the extent and number of observed magnetic field rotations in magnetic hodograms. This method avoids the introduction of further assumptions on the 3D magnetic configuration as required by more advanced fitting techniques.

As can be seen from the hodograms (Figure~\ref{fig:hodograms}), the dominant rotation is in the $B_N$ component at both spacecraft, with $B_R$ also changing sign at PSP. Overall, there is a mostly negative $B_R$ component at both PSP and STEREO-A, which could be due to the spacecraft crossing below the flux rope axis, i.e., that the flux rope axis passed north of both PSP and STEREO-A. As described in Section~\ref{sec:remote}, given the fact that we only have one viewpoint from the limb of the CME, its initial direction cannot be deciphered. Although oriented slightly differently, overall we find a SWN right-handed flux rope with Fr classification at both PSP and STEREO-A. {By SWN flux rope we understand a low inclination right-handed flux rope, which has a primary rotation from south to north, pointing to the west (following the classifications by \citet{Bothmer1998} and \citet{Mulligan1998}).} This implies an approximate flux rope axis orientation of $\theta \sim 0^{\circ}$ and $\phi \sim 90^{\circ}$ in RTN coordinates.

\subsection{Background Solar Wind Conditions}
\label{subsec:backgrond_solar_wind_conditions}

To contextualise the CME propagation with respect to the conditions of interplanetary space, we model the solar wind using the ENLIL heliospheric model \citep{Odstrcil2003}, running at the NASA Community Coordinated Modeling Center (CCMC) and available for runs on request (\url{https://ccmc.gsfc.nasa.gov/requests/SH/E28/enlil_options.php}). 
We perform one low resolution simulation (i.e.\ 384 grid cells in the radial direction, 30 grid cells in the latitudinal direction, and 90 grid cells in the longitudinal direction) spanning between 0.1 and 2.1~au in the radial direction, 
and covering $\pm 60^\circ$ in latitude and $\pm 180^\circ$ in longitude.
We run ENLIL using as inner boundary conditions coronal solutions obtained by the Wang-Sheeley-Arge (WSA) semi-empirical coronal model \citep{Arge2000, Arge2004} using a daily-updated map generated by the Global Oscillation Network Group (GONG) network on October 10, 2019.
The simulation results can be freely accessed through the CCMC webpage (\url{https://ccmc.gsfc.nasa.gov/database_SH/Camilla_Scolini_062520_SH_2.php}).
Selected snapshots of the modeled solar wind speed close to the heliographic equatorial plane are provided in Figure~\ref{fig:obs_mod_summary}(d), (e), (f).

\begin{figure}
\centering
{\includegraphics[width=0.8\hsize]{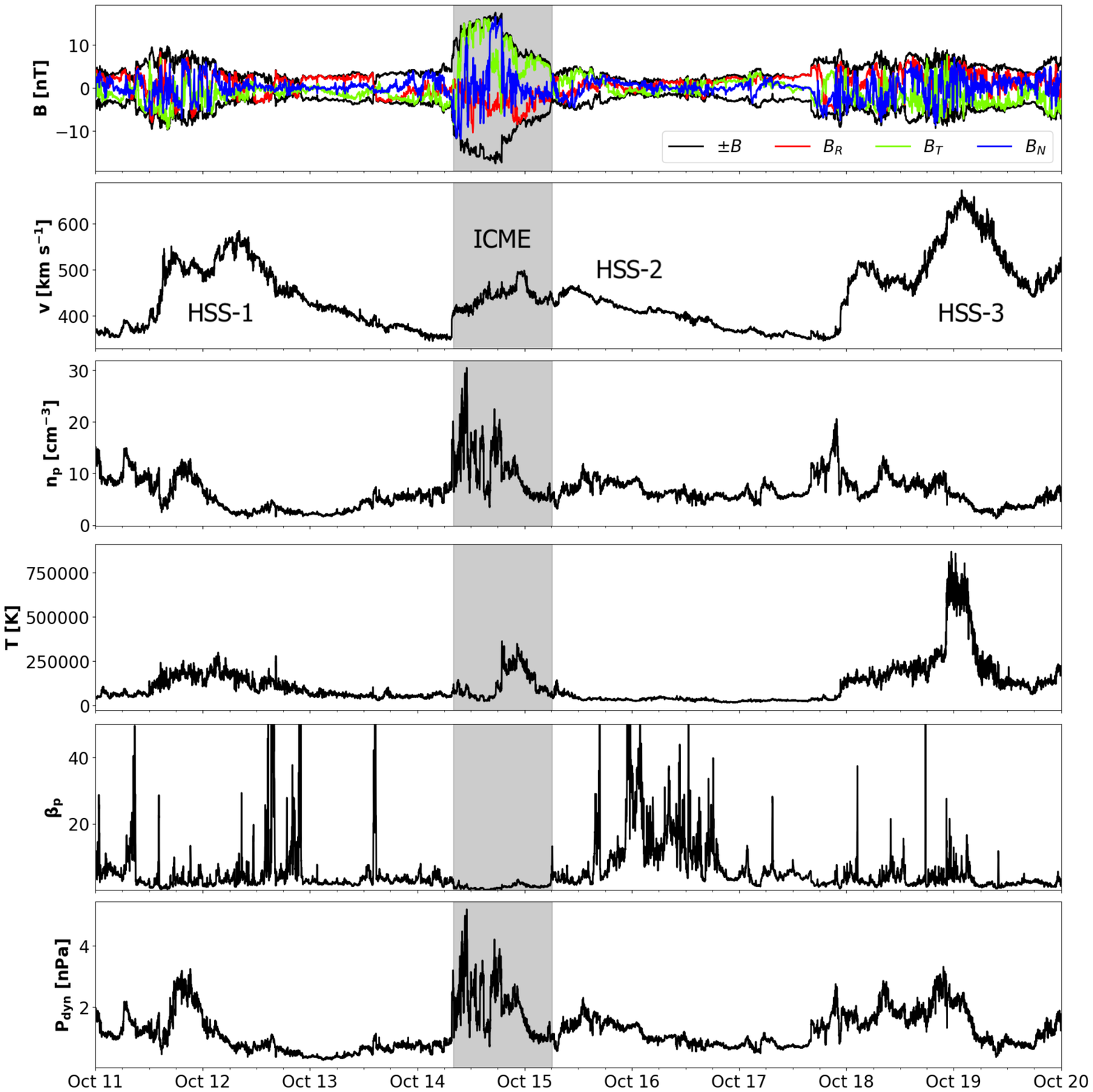}}
\caption{
Magnetic field and plasma data at STEREO-A for the period between October 11, and October 19, 2019. 
The top panel shows the magnetic field data in RTN coordinates, using the same colors as in Figure~\ref{fig:CMEatSTA}.  
The following panels show the speed, proton number density, temperature, proton $\beta$, and dynamic pressure, respectively.  
The grey shaded area marks the passage of the CME, from the detection of the preceding shock to the end of the CME signature.
In the second panel, the signatures of HSS-1, HSS-2, and HSS-3 are also marked.
} 
\label{fig:CMEatSTA_supplemental}
\end{figure}

These ENLIL simulations of the solar wind show the presence of three HSSs near PSP and STEREO-A around the same time as the CME. In Figure~\ref{fig:obs_mod_summary}(d), we have labeled these HSSs according to their arrival time at PSP and STEREO-A. Around the time that the CME was ejected from the Sun, HSS-1 had just passed by PSP and STEREO-A (Figure~\ref{fig:obs_mod_summary}(d)), while HSS-2 arrived at PSP and STEREO-A around the same time as the CME (Figure~\ref{fig:obs_mod_summary}(e), (f)). Finally, HSS-3 arrived about 5 days after HSS-2 at PSP and STEREO-A (arrival not shown in Figure~\ref{fig:obs_mod_summary}). Given the position and arrival time of the CME of interest with respect to the aforementioned HSSs, the CME likely originated from the white region in between the two coronal holes in Figure~\ref{fig:obs_mod_summary}(b), which are most likely the sources of HSS-1 (western coronal hole) and HSS-2 (central coronal hole). 

Magnetic field and solar wind plasma data at STEREO-A confirm that the CME is preceded by and closely followed by HSSs, with HSS-1 arriving a few days before the CME, HSS-2 arriving right on the tail of the CME, and HSS-3 a few days after the CME (see Figure~\ref{fig:CMEatSTA_supplemental}). These in situ measurements indicate an overall qualitative match between the in situ data and ENLIL simulations at this time, and also show that HSS-1 and HSS-3 are associated with CIRs. With regards to HSS-2, a clear association with a CIR is not present as there is no clear shock or increase in temperature, only a small density increase is observed.

Comparing the timing of the HSS' arrivals between the ENLIL simulations and the in situ measurements at STEREO-A, we find that the ENLIL model predicts the arrival time of HSS-1 and HSS-2 $\sim 1.5$~days too early, while the arrival time of HSS-3 is predicted around one day later than actually observed. These sub-optimal arrival time predictions are likely the result of the specific morphology of the coronal holes from where each stream originates, and of the fact that the photospheric magnetic field data used as input for the simulation are only updated in regions on the solar disk that are seen as front-sided from Earth (i.e.\ opposite to the direction of propagation of HSS-2 and HSS-3).

Overall, the qualitative agreement between the ENLIL simulations and the in situ data at STEREO-A confirms that HSS-2 arrived right on the tail of the CME, and interacted with the CME as it propagated from the Sun to PSP and to STEREO-A. Due to the increased speed and dynamic pressure associated with HSS-2 compared to nominal solar wind conditions, the nearly linear CME speed increase from the Sun to 1~au and the compression signatures observed in the CME (including the increased CME transit speed between the Sun and STEREO-A as compared to between the Sun and PSP), the shorter CME duration at STEREO-A compared to PSP, and the increasing speed profile in the ME at STEREO-A) are most likely caused by this high speed stream. HSS-2 thus likely accelerated and altered the CME before it even reached PSP, possibly causing the reconnection features in the ME tail as seen in the solar wind plasma data along with disconnection from the Sun as evidenced by the entirely unidirectional suprathermal electrons in the ME tail at STEREO-A. As mentioned at the end of Section~\ref{subsec:shock}, {the fact that} this initially very slow CME drives a shock at STEREO-A indicates that the interaction between the CME and HSS-2 must have occurred for a relatively long time, $\sim 2.5$~days, by the time it reached STEREO-A, which is consistent with the origin of the CME just west of the coronal hole associated with HSS-2.

\section{Conclusions}\label{sec:conclusion}

In this paper, we have detailed observations of the first CME detected in simultaneous conjunction at PSP and STEREO-A. Despite it being a stealth CME and described as a ``very poor" event in the CDAW catalog, there are clear in situ signatures of the CME at PSP and STEREO-A with typical CME magnetic field increases and CME substructures. Given that most conjunction CMEs studied in detail in the past have been halo CMEs with fast speeds, this slow CME presents a relatively unique opportunity to study its structure and evolution, specifically as a result of its interaction with an overtaking HSS.

From in situ data at PSP and STEREO-A, we found that only about half of the duration of the CME ME is made up of a flux rope, while the other half is a long tail region with no flux rope signatures at both spacecraft. Furthermore, there are clear compression signatures in the CME, including the increasing speed profile in the ME at STEREO-A, the shorter CME duration at STEREO-A (i.e., at a larger heliocentric distance) than at PSP, and the estimated faster CME speed at STEREO-A than at PSP. Based on the in situ data at STEREO-A and ENLIL simulations of the solar wind, we found that the aforementioned distinctive features were caused by a solar wind HSS accelerating and overtaking this slow CME. Despite alterations, the CME magnetic field is highly correlated between PSP and STEREO-A, i.e. the same distortions exist at both. We estimated a minimum interaction time of $\sim$2.5 days between the CME and the HSS by the time the CME reached STEREO-A, based on the fact that the CME was able to drive a shock at STEREO-A. This lengthy interaction likely enabled the almost linear speed increase of this initially very slow CME to speeds well beyond that of the slow solar wind ahead of it. In addition, it led to the formation of a shock wave ahead of the CME at STEREO-A and potentially at PSP, a highly unexpected result for a CME with an initial speed below 200~km\,s$^{-1}$. Such a shock may be able to accelerate particles to suprathermal energies as it propagates further and contribute to the formation of a reservoir in the inner heliosphere. 

The condition observed of persistent, steady acceleration of a small CME is relatively unusual. More typically, small CMEs accelerate inside of 1~au, and then reach a steady speed as they move with the ambient solar wind. In the current case, the CME moves out in a region where fast solar wind overtakes slower wind ahead of it. The fast wind overtaking slow wind generally increases the speed of slower wind beyond the stream interface. Since the CME is embedded within this interaction region, the CME itself is accelerated by the faster flow behind it, which helps explain the persistent acceleration observed. Furthermore, steady acceleration of the CME requires coherence in the CME magnetic structure, which is also evidenced by the high magnetic field correlation between the PSP and STEREO-A measurements. Future observations of slow CMEs in conjunction between PSP and other spacecraft will shed more light on the physical processes affecting the propagation of such CMEs.

\section{Data availability} All the data analyzed in this study are publicly available. PSP and STEREO data are available on the Space Physics Data Facility's Coordinated Data Analysis Web ({\url{https://cdaweb.sci.gsfc.nasa.gov}}).

\acknowledgments
R. M. W. acknowledges support from NASA grant 80NSSC19K0914. N. L. acknowledges support from NASA grants 80NSSC20K0700 and 80NSSC17K0009. 
C. S. acknowledges support from the Research Foundation -- Flanders (FWO, PhD fellowship no. 1S42817N), and from the NASA Living With a Star Jack Eddy Postdoctoral Fellowship Program, administered by UCAR's Cooperative Programs for the Advancement of Earth System Science (CPAESS) under award no. NNX16AK22G. R. M. W., N. L., and A. B. G. were partially supported by the NASA STEREO grant 80NSSC20K0431.

The FIELDS experiment on the Parker Solar Probe spacecraft was designed and developed under NASA contract NNN06AA01C. Simulation results have been provided by the Community Coordinated Modeling Center (CCMC) at NASA Goddard Space Flight Center through their public Runs on Request system (\url{http://ccmc.gsfc.nasa.gov}). The CCMC is a multi-agency partnership between NASA, AFMC, AFOSR, AFRL, AFWA, NOAA, NSF and ONR. The ENLIL model was developed by D. Odstr\v{c}il at the University of Colorado at Boulder.

\bibliography{Refs}{}
\bibliographystyle{aasjournal}



\end{document}